\documentclass[conference]{IEEEtran}
\IEEEoverridecommandlockouts
\usepackage{cite}
\usepackage{amsmath,amssymb,amsfonts}
\usepackage{algorithmic}
\usepackage{graphicx}
\usepackage{subfig}
\usepackage{textcomp}
\usepackage{xcolor}
\def\BibTeX{{\rm B\kern-.05em{\sc i\kern-.025em b}\kern-.08em
    T\kern-.1667em\lower.7ex\hbox{E}\kern-.125emX}}
\begin{document}

\title{Deep convolutional neural networks for cyclic sensor data\\

}
\author{\IEEEauthorblockN{1\textsuperscript{st} Payman Goodarzi}
\IEEEauthorblockA{\textit{Lab for Measurement Technology} \\
\textit{Saarland University}\\
Saarbruecken, Germany \\
p.goodarzi@lmt.uni-saarland.de}
\and
\IEEEauthorblockN{2\textsuperscript{rd} Yannick Robin}
\IEEEauthorblockA{\textit{Lab for Measurement Technology} \\
\textit{Saarland University}\\
Saarbruecken, Germany \\
y.robin@lmt.uni-saarland.de}
\and
\IEEEauthorblockN{3\textsuperscript{th} Andreas Schütze}
\IEEEauthorblockA{\textit{Lab for Measurement Technology} \\
\textit{Saarland University}\\
Saarbruecken, Germany \\
schuetze@lmt.uni-saarland.de}
\and
\IEEEauthorblockN{4\textsuperscript{th} Tizian Schneider}
\IEEEauthorblockA{\textit{Lab for Measurement Technology} \\
\textit{Saarland University}\\
Saarbruecken, Germany \\
t.schneider@lmt.uni-saarland.de}
}

\maketitle

\begin{abstract}
Predictive maintenance plays a critical role in ensuring the uninterrupted operation of industrial systems and mitigating the potential risks associated with system failures. This study focuses on sensor-based condition monitoring and explores the application of deep learning techniques using a hydraulic system testbed dataset. Our investigation involves comparing the performance of three models: a baseline model employing conventional methods, a single CNN model with early sensor fusion, and a two-lane CNN model (2L-CNN) with late sensor fusion. The baseline model achieves an impressive test error rate of 1\% by employing late sensor fusion, where feature extraction is performed individually for each sensor. However, the CNN model encounters challenges due to the diverse sensor characteristics, resulting in an error rate of 20.5\%. To further investigate this issue, we conduct separate training for each sensor and observe variations in accuracy. Additionally, we evaluate the performance of the 2L-CNN model, which demonstrates significant improvement by reducing the error rate by 33\% when considering the combination of the least and most optimal sensors. This study underscores the importance of effectively addressing the complexities posed by multi-sensor systems in sensor-based condition monitoring.
\end{abstract}


\section{Introduction}
Industrial systems and factories operate continuously, necessitating uninterrupted performance to avoid process downtime, significant financial losses, and potential safety hazards. To mitigate these risks, companies employ various maintenance approaches, including corrective maintenance, preventive maintenance, and predictive maintenance. Predictive maintenance (PdM) heavily relies on monitored signals from diverse sensors, with machine learning methods playing a pivotal role in data-driven PdM. These methods can be categorized into two groups: conventional approaches and deep learning techniques. Conventional methods involve preprocessing, feature extraction (FE), feature selection (FS), and the subsequent application of classification or regression algorithms \cite{Schneider2019}, commonly referred to as FESC/FESR in this study. In contrast, modern deep neural networks have demonstrated exceptional performance across various applications, including PdM \cite{Namuduri2020,Zhao2019,Magar2021}.

In line with these advancements, gas mixture measurement has emerged as a promising application for deep neural networks in recent research. Notably, Robin et al. \cite{Robin2021} introduced a convolutional neural network (CNN) specifically designed for indoor air quality monitoring (TCOCNN), accurately predicting volatile organic compounds using temperature-cycled operation sensors. The proposed method surpassed existing data evaluation techniques, underscoring the effectiveness of CNNs in this domain. It is worth noting that the signals utilized in their study bear resemblance to the typical data encountered in condition monitoring and predictive maintenance applications, i.e. multiple sensors with periodic or cyclic data.

Motivated by the aforementioned findings, the objective of our study is to compare our previously published method, TCOCNN, with a benchmark method in a different application context. To achieve this, we utilize a publicly available dataset from a hydraulic system testbed \cite{Helwig2015}. Recent research has applied various deep learning techniques to the dataset under investigation \cite{Prakash2023, Pillai2022, Zhang2022, Berghout2021, Ma2021}. Prakash et al. \cite{Prakash2023} employed a 1D CNN model to analyze the pressure difference between two pressure sensors. Huang et al. \cite{Huang2022} took a parallel approach by utilizing multiple independent convolutional neural networks to extract features from individual sensors. Furthermore, Berghout et al. \cite{Berghout2021} introduced a novel neural network model specifically designed to process the extracted features. In a distinct approach, Zhang et al. \cite{Zhang2022} demonstrated the application of a Transformer model with self-attention, originally trained on natural language, to the task of sensor fusion. Collectively, these studies contribute to the exploration of diverse methodologies for analyzing sensor data and extracting meaningful insights. The primary goal of our comprehensive evaluation is to assess the performance of the air quality model when applied to the field of condition monitoring. In doing so, we aim to address potential challenges and difficulties associated with multiple sensors of different types, thereby providing valuable insights for future research in this area. The remaining structure of this paper is outlined as follows: Section \ref{sec:MM} describes the materials and methods, including details about the dataset and the utilized convolutional neural network. Section \ref{sec:Res} reports the results, and finally, Section \ref{sec:Con} presents the conclusions derived from this study.

\section{Materials and methods}
\label{sec:MM}
\subsection{Dataset}

\begin{figure}[htbp]
\centering
\subfloat[][]{\includegraphics[width=1.7in]{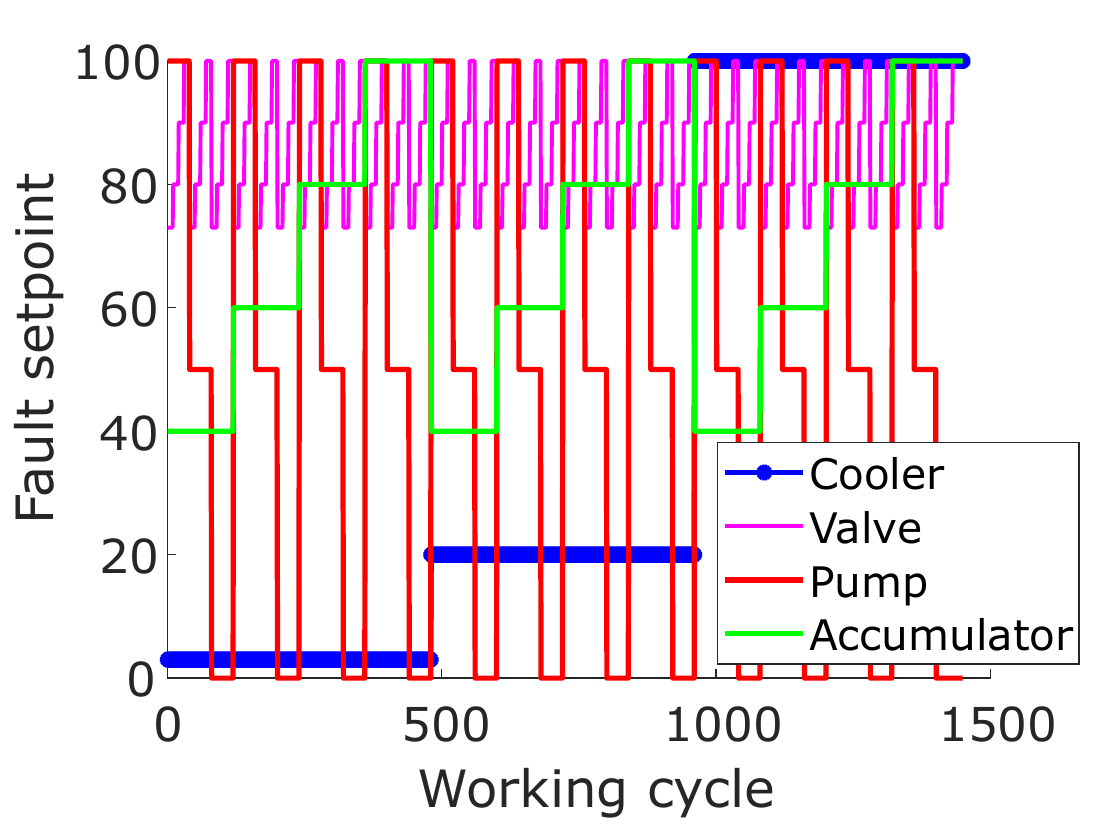}\label{fig_hs1}}
\hfil
\subfloat[][]{\includegraphics[width=1.7in]{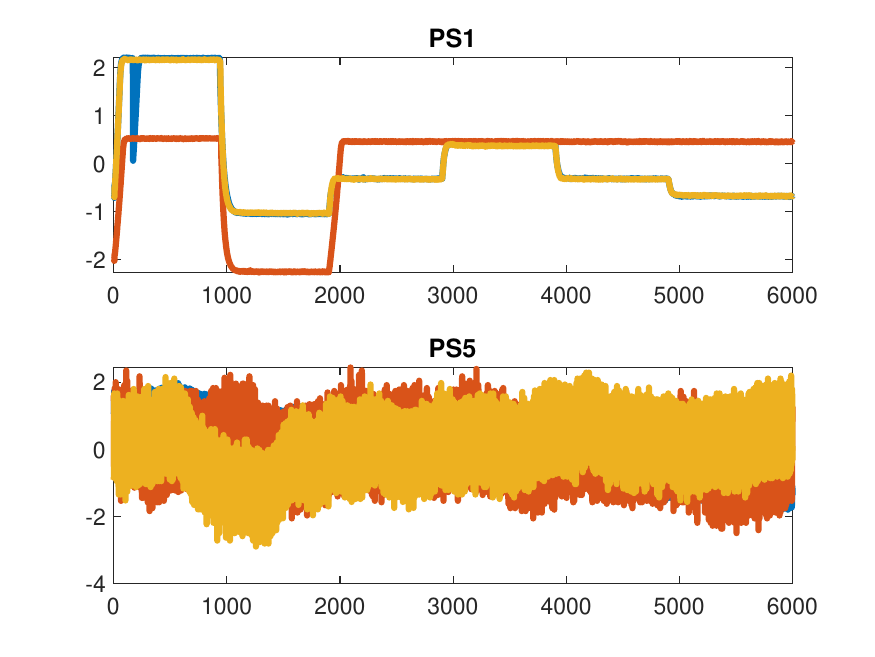}\label{fig_hs2}}
\caption{ZeMA dataset control variables (a), and three cycles of two selected sensors (b).}
\label{fig_hs}
\end{figure}

This study utilizes a dataset that captures the behavior of a hydraulic system (HS) testbed, which has been specifically designed to simulate various common faults encountered in such systems \cite{Helwig2015}. The ZeMA\footnote{Zentrum für Mechatronik und Automatisierungstechnik gemeinnützige GmbH} dataset includes simulated faults such as decreased cooler performance, main valve switching degradation, internal pump leakage, and accumulator pre-charge pressure reduction with the control system enabling independent adjustment of each fault condition. Fig. \ref{fig_hs1} provides an illustration of the conditions of the cooler, valve, pump, and accumulator within the dataset, which consists of recordings from 17 sensors over a constant operating cycle lasting 60 seconds as would be typical, e.g. for a hydraulic press operation. These sensors measure process values, including pressure (PS1 - PS6), flow (FS1, FS2), temperature (TS1 - TS5), electrical power (EPS1), and vibration (VS1). Additionally, the dataset includes three virtual sensors, namely cooling efficiency (CE), cooling power (CP), and system efficiency (SE). These virtual sensors are calculated using a physical model that combines various measured values. The dataset comprises sensors of various types, and their sampling rates vary based on the measured parameter. The sampling frequencies range from 1 to 100 Hz, resulting in observations with 60 to 6000 data samples per sensor per cycle. Fig. \ref{fig_hs2} effectively showcases the distinct characteristics exhibited by two sensors through three cycles, highlighting the multimodality of the sensor data \cite{Baltrusaitis2019}. In the context of the present study, the target is to predict the condition of the accumulator. Specifically, the model aims to classify the hydraulic accumulator pre-charge pressure into categories, i.e., "optimal pressure," "lightly reduced pressure," "severely reduced pressure," and "close to total failure."

\subsection{Algorithms}
Conventional ML:
The baseline model for this study is selected from \cite{Schneider2018}, where an AutoML toolbox is employed to analyze the dataset. The AutoML toolbox explores various combinations of FESC methods to identify the most effective approach. In this study, we utilize the method identified by the toolbox, which achieved the highest cross-validation accuracy. The selected method involves extracting statistical moments (mean, standard deviation, skewness, and kurtosis) from the raw data as the features. Pearson correlation is used as the feature selector, and linear discriminant analysis with Mahalanobis distance serves as the classifier.

\begin{figure}[htbp]
\centering
\centerline{\includegraphics[width=3in]{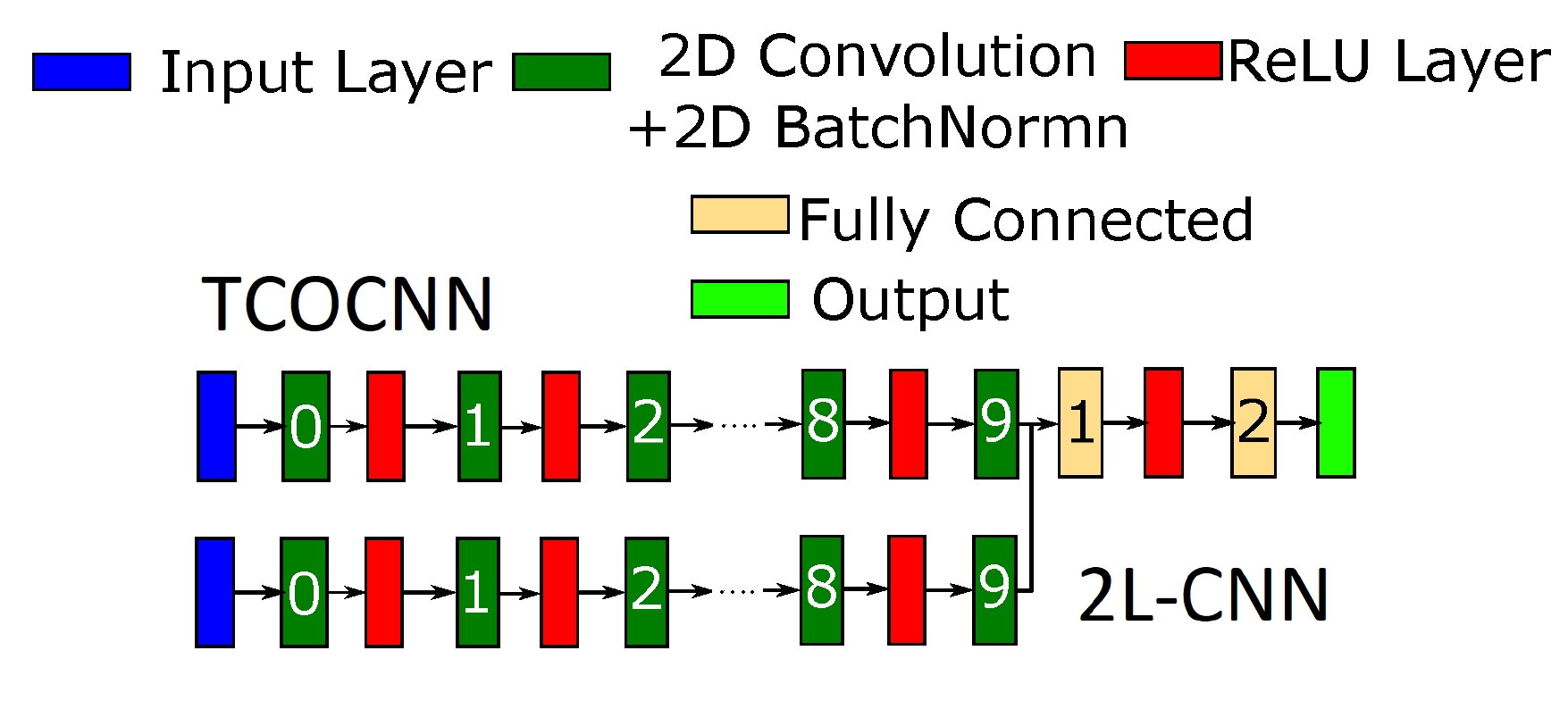}}
\caption{Architecture of the TCOCNN and 2L-CNN.}
\label{figNet}
\end{figure}

Deep learning methods:
Deep learning techniques were employed as the second approach to construct the models. Specifically, two CNN models were utilized: TCOCNN and 2L-CNN, as depicted in Figure \ref{figNet}. TCOCNN is a deep network comprising 10 convolutional layers, and we also evaluated the late fusion version of TCOCNN, which has demonstrated its effectiveness in various multimodal systems \cite{Joze2020, Huang2022}. In the 2L-CNN model, each sensor was assigned its own set of convolutional filters. The resulting feature maps from each convolutional lane were concatenated, and a fully connected layer was then applied.

In a previous study \cite{Robin2021}, the employed model was utilized to predict multiple gas concentrations, such as acetone, ethanol, formaldehyde, toluene, the total concentration of all volatile organic compounds (VOCsum), as well as the inorganic gases carbon monoxide and hydrogen. However, in the present study, a different approach is adopted as it focuses on a classification task. Consequently, the last fully connected layer and output function differ from those used in \cite{Robin2021}. To determine the network's hyperparameters, a hyperparameter (HP) tuning process is employed. Due to computational resource limitations, the HP tuning involves conducting 50 iterations within a predefined search space for the HPs, as detailed in Table \ref{tab1}.

\begin{table*}[t!]
\caption{The hyperparameter ranges for the CNN.}
\centering
\footnotesize
\begin{tabular}{|p{2.5cm}|p{2.5cm}|p{2.5cm}|p{2.5cm}|p{2.5cm}|p{2.5cm}|}
\hline
\textbf{Initial Learning Rate (Log Scale)} & \textbf{Number of Filters (First Two Layers)} & \textbf{Kernel Size (First Two Layers)} & \textbf{Stride Size (First Layer)} & \textbf{Dropout Number of Neurons (FC)} & \textbf{Number of Neurons (FC)} \\
\hline
$1 \times 10^{-7} - 1 \times 10^{-4}$ & 10-100 & 100-300 & 100-175 & 30-50\% & 500-2500 \\
\hline
\end{tabular}
\label{tab1}
\end{table*}

To address the issue of multiple sensors having different sampling rates and, thus, lengths, a preprocessing step is employed where all sensors are upsampled to 6000 data points, ensuring the same number of raw data for each sensor. These upsampled sensor readings are then combined to form a 2D matrix, enabling the application of a 2D convolutional network. Following the preprocessing step, a random split is performed on the dataset. The data is divided into three subsets: 70\% of the data is allocated for training, 10\% for validation, and 20\% for testing. This partitioning ensures that the model is trained on a substantial portion of the data while having separate datasets for validation and testing. The HP tuning is then exclusively performed on the training data, enabling the identification of optimal HPs based on minimizing the validation loss. This ensures the model is fine-tuned and optimized for performance on unseen data.

\section{Results}
\label{sec:Res}
The results obtained from the baseline model, which utilizes conventional methods, showcased excellent performance for the defined task. When all 17 sensors were used as input, the model achieved a test error rate of 1\%. The conventional method applies feature extraction to each sensor individually, thereby avoiding any challenges associated with multiple variant sensors.

\begin{figure}[htbp]
\centering
\subfloat[]{\includegraphics[width=1.6in]{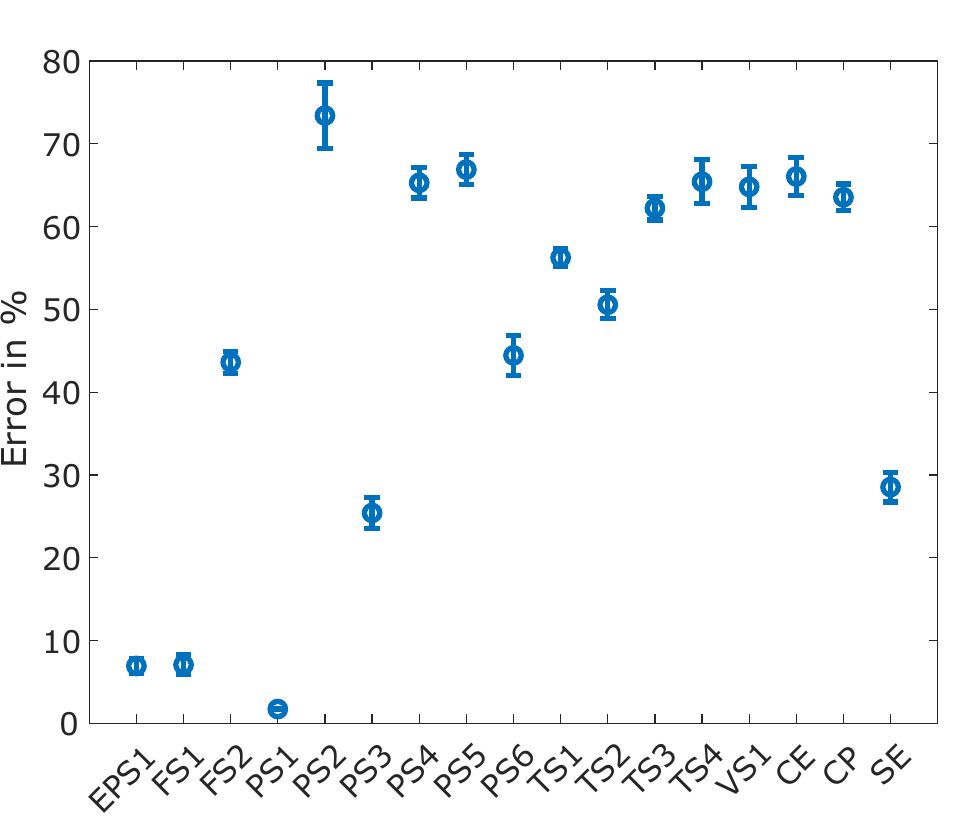}
\label{fig_results2}}
\hfil
\subfloat[]{\includegraphics[width=1.6in]{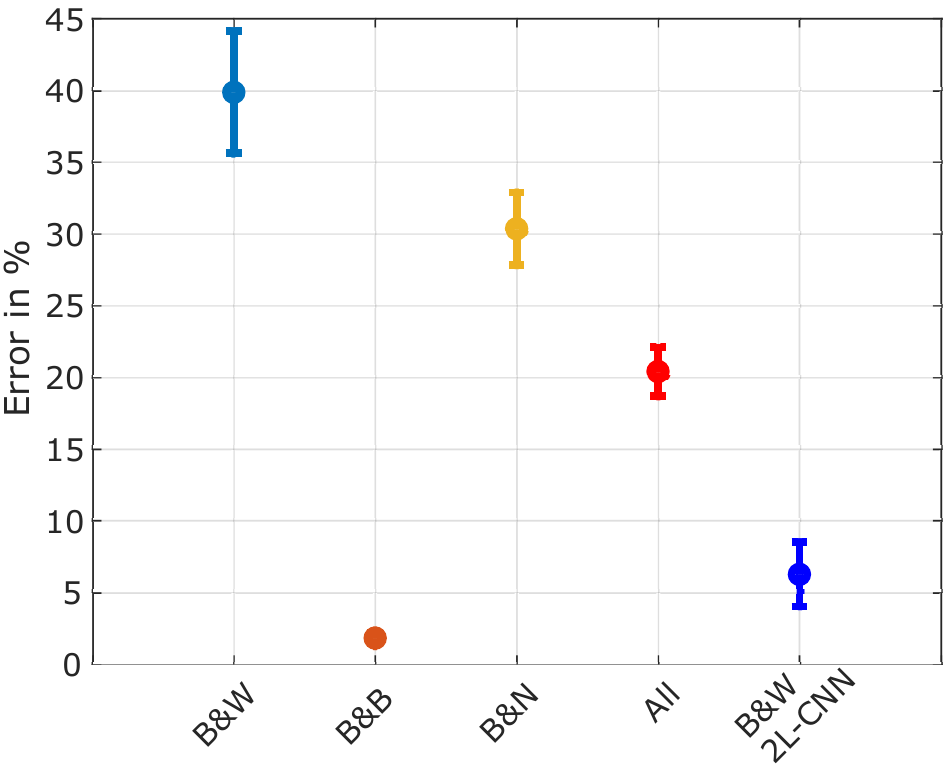}
\label{fig_results1}}
\caption{Test error when only single sensors are used (a), and when combinations of sensors are used (b). Each HP tuning is repeated 5 times, and the error bars represent the range for the standard deviation of the results.}
\label{fig_sim2}
\end{figure}

In contrast, the CNN model incorporating all 17 sensors and the mentioned preprocessing steps exhibited lower performance, achieving an error rate of 20.5\%. This outcome highlighted the challenges arising from the multimodal characteristics of the sensors. To highlight this issue, we trained the network separately for each sensor and observed varying error rates. Fig. \ref{fig_results2} displays the error results, with PS1 achieving the best error rate of 1.7\% and PS2 obtaining the highest error of 73.4\%.

To further investigate the impact of dissimilar sensors on the deep neural network model, we conducted a series of tests using different combinations of sensors. In order to keep the experiments manageable, we focused on two sensors and trained the network using all possible combinations involving the best sensor (PS1) and the remaining sensors. The results are illustrated in Figure 1, where the most notable findings are highlighted. The worst combination, involving PS1 and PS5, exhibited a significantly higher error rate of 39.8\% (B\&W). Moreover, when comparing the performance of using the best sensor alone to using it twice (B\&B), we observed that increasing the input size (2x6000) did not have a significant impact on the results. However, introducing a second sensor with uniform noise (B\&N) as an input resulted in a substantial increase in the error rate, reaching 30.3\%.

Fig. \ref{fig_results1} provides an illustration of the result achieved by this new model, denoted as "2L-CNN," using the best and worst sensors as input. The new model demonstrated improved performance compared to the normal single-lane network, achieving error rates of 39.8\% and 6.5\% respectively. This finding provides additional support for the importance of effectively addressing the challenges associated with dissimilar sensors and multimodal learning in order to maintain and improve performance.

\section{Conclusion}
\label{sec:Con}
In conclusion, this study aimed to assess the performance of a network originally designed for temperature-cycled gas sensors in a different application, namely the fault classification of a hydraulic system, and compare it with conventional methods. The conventional baseline model demonstrated impressive performance in handling the 17 diverse sensors, achieving an outstanding error rate of 1\%. This method utilizes independent late data fusion after extracting features individually from each sensor.

In contrast, the CNN model that incorporated all 17 sensors and utilized preprocessing to achieve the same raw data length demonstrated significantly lower performance, resulting in an error rate of 20.5\%. This outcome highlighted the challenges arising from the dissimilar characteristics of the sensors, which hindered the network's ability to effectively handle them. Notably, training the network separately for each sensor revealed substantial variations in error rates, ranging from 1.7\% for PS1 to 73.4\% for PS2.

To emphasize the importance of pertinent input data, we performed experiments using various sensor combinations. These tests provided clear evidence that incorporating irrelevant sensors in the input data significantly compromised the results. Additionally, we evaluated the performance of a 2L-CNN model that utilizes late-sensor fusion, which proved to be a viable strategy for addressing the issues caused by irrelevant sensors. This discovery underscores the crucial significance of meticulously choosing and prioritizing relevant sensors to enhance the model's performance.

Overall, this study underscores the challenges posed by the multimodal nature of sensor data. It emphasizes the significance of effectively addressing these challenges to unlock the full potential of sensor-based applications and enhance their overall performance. Future research endeavors can focus on exploring advanced techniques to overcome these obstacles and further enhance the accuracy of fault detection models.

\bibliographystyle{IEEEtran}
\bibliography{sensor23}

\begin{thebibliography}{10}
\providecommand{\url}[1]{#1}
\csname url@samestyle\endcsname
\providecommand{\newblock}{\relax}
\providecommand{\bibinfo}[2]{#2}
\providecommand{\BIBentrySTDinterwordspacing}{\spaceskip=0pt\relax}
\providecommand{\BIBentryALTinterwordstretchfactor}{4}
\providecommand{\BIBentryALTinterwordspacing}{\spaceskip=\fontdimen2\font plus
\BIBentryALTinterwordstretchfactor\fontdimen3\font minus
  \fontdimen4\font\relax}
\providecommand{\BIBforeignlanguage}[2]{{%
\expandafter\ifx\csname l@#1\endcsname\relax
\typeout{** WARNING: IEEEtran.bst: No hyphenation pattern has been}%
\typeout{** loaded for the language `#1'. Using the pattern for}%
\typeout{** the default language instead.}%
\else
\language=\csname l@#1\endcsname
\fi
#2}}
\providecommand{\BIBdecl}{\relax}
\BIBdecl

\bibitem{Schneider2019}
T.~Schneider, S.~Klein, and A.~Schütze, ``Machine learning in industrial
  measurement technology for detection of known and unknown faults of equipment
  and sensors,'' \emph{Technisches Messen}, vol.~86, pp. 706--718, 11 2019.

\bibitem{Namuduri2020}
S.~Namuduri, B.~N. Narayanan, V.~S.~P. Davuluru, L.~Burton, and S.~Bhansali,
  ``Review—deep learning methods for sensor based predictive maintenance and
  future perspectives for electrochemical sensors,'' \emph{Journal of The
  Electrochemical Society}, vol. 167, p. 037552, 2 2020.

\bibitem{Zhao2019}
\BIBentryALTinterwordspacing
R.~Zhao, R.~Yan, Z.~Chen, K.~Mao, P.~Wang, and R.~X. Gao, ``Deep learning and
  its applications to machine health monitoring,'' \emph{Mechanical Systems and
  Signal Processing}, vol. 115, pp. 213--237, 2019. [Online]. Available:
  \url{https://www.sciencedirect.com/science/article/pii/S0888327018303108}
\BIBentrySTDinterwordspacing

\bibitem{Magar2021}
R.~Magar, L.~Ghule, J.~Li, Y.~Zhao, and A.~B. Farimani, ``Faultnet: A deep
  convolutional neural network for bearing fault classification,'' \emph{IEEE
  Access}, vol.~9, pp. 25\,189--25\,199, 2021.

\bibitem{Robin2021}
Y.~Robin, J.~Amann, T.~Baur, P.~Goodarzi, C.~Schultealbert, T.~Schneider, and
  A.~Schütze, ``High-performance voc quantification for iaq monitoring using
  advanced sensor systems and deep learning,'' \emph{Atmosphere}, vol.~12, p.
  1487, 11 2021.

\bibitem{Helwig2015}
N.~Helwig, E.~Pignanelli, and A.~Schutze, ``Condition monitoring of a complex
  hydraulic system using multivariate statistics.''\hskip 1em plus 0.5em minus
  0.4em\relax IEEE International Instrumentation and Measurement Technology
  Conference (I2MTC) Proceedings, 5 2015, pp. 210--215.

\bibitem{Prakash2023}
J.~Prakash, S.~Singh, A.~Miglani, and P.~K. Kankar, ``Pressure signal-based
  analysis of anomalies in switching behavior of a two-way directional control
  valve,'' \emph{ASME Open Journal of Engineering}, vol.~2, 1 2023.

\bibitem{Pillai2022}
S.~Pillai and P.~Vadakkepat, ``Deep learning for machine health prognostics
  using kernel-based feature transformation,'' \emph{Journal of Intelligent
  Manufacturing}, vol.~33, pp. 1665--1680, 8 2022.

\bibitem{Zhang2022}
Z.~Zhang, M.~Farnsworth, B.~Song, D.~Tiwari, and A.~Tiwari, ``Deep transfer
  learning with self-attention for industry sensor fusion tasks,'' \emph{IEEE
  Sensors Journal}, vol.~22, pp. 15\,235--15\,247, 8 2022.

\bibitem{Berghout2021}
T.~Berghout, M.~Benbouzid, S.~M. Muyeen, T.~Bentrcia, and L.-H. Mouss,
  ``Auto-nahl: A neural network approach for condition-based maintenance of
  complex industrial systems,'' \emph{IEEE Access}, vol.~9, pp.
  152\,829--152\,840, 2021.

\bibitem{Ma2021}
X.~Ma, P.~Wang, B.~Zhang, and M.~Sun, ``A multirate sensor information fusion
  strategy for multitask fault diagnosis based on convolutional neural
  network,'' \emph{Journal of Sensors}, vol. 2021, pp. 1--17, 6 2021.

\bibitem{Huang2022}
K.~Huang, S.~Wu, F.~Li, C.~Yang, and W.~Gui, ``Fault diagnosis of hydraulic
  systems based on deep learning model with multirate data samples,''
  \emph{IEEE Transactions on Neural Networks and Learning Systems}, vol.~33,
  pp. 6789--6801, 11 2022.

\bibitem{Baltrusaitis2019}
T.~Baltrusaitis, C.~Ahuja, and L.-P. Morency, ``Multimodal machine learning: A
  survey and taxonomy,'' \emph{IEEE Transactions on Pattern Analysis and
  Machine Intelligence}, vol.~41, pp. 423--443, 2 2019.

\bibitem{Schneider2018}
\BIBentryALTinterwordspacing
T.~Schneider, N.~Helwig, and A.~Schütze, ``Industrial condition monitoring
  with smart sensors using automated feature extraction and selection,''
  \emph{Measurement Science and Technology}, vol.~29, p. 94002, 8 2018.
  [Online]. Available: \url{https://doi.org/10.1088/1361-6501/aad1d4}
\BIBentrySTDinterwordspacing

\bibitem{Joze2020}
H.~R.~V. Joze, A.~Shaban, M.~L. Iuzzolino, and K.~Koishida, ``Mmtm: Multimodal
  transfer module for cnn fusion.''\hskip 1em plus 0.5em minus 0.4em\relax
  Proceedings of the IEEE/CVF Conference on Computer Vision and Pattern
  Recognition, 2020, pp. 13\,289--13\,299.

\end{thebibliography}
\vspace{12pt}

\end{document}